\documentclass[publisher]{aa}
 \bibpunct{(}{)}{;}{a}{}{,}
 \usepackage[varg]{txfonts}
 \usepackage{color}
 \usepackage{url}

\begin{document}
\title{Number density structures in the inner heliosphere}
\author{D. Stansby \and T. S. Horbury}
\institute{Department of Physics, Imperial College London, London, SW7 2AZ, United Kingdom \email{david.stansby14@imperial.ac.uk}}


\abstract
{}
{The origins and generation mechanisms of the slow solar wind are still unclear. Part of the slow solar wind is populated by ``number density structures'', discrete patches of increased number density that are frozen in to and move with the bulk solar wind. In this paper we aimed to provide the first in-situ statistical study of number density structures in the inner heliosphere.}
{We reprocessed in-situ ion distribution functions measured by Helios in the inner heliosphere to provide a new reliable set of proton plasma moments for the entire mission. From this new data set we looked for number density structures measured within 0.5 AU of the Sun and studied their properties.}
{We identified 140 discrete areas of enhanced number density. The structures occurred exclusively in the slow solar wind and spanned a wide range of length scales from 50 Mm to 2000 Mm, which includes smaller scales than have been previously observed. They were also consistently denser and hotter that the surrounding plasma, but had lower magnetic field strengths, and therefore remained in pressure balance.}
{Our observations show that these structures are present in the slow solar wind at a wide range of scales, some of which are too small to be detected by remote sensing instruments. These structures are rare, accounting for only 1\% of the slow solar wind measured by Helios, and are not a significant contribution to the mass flux of the solar wind.}

\maketitle

\section{Introduction}
The solar wind is traditionally viewed as bimodal, consisting of `fast' ($ > \sim$ 500 km/s) and `slow' ($< \sim$ 500 km/s) components. It is well established that fast solar wind originates from coronal holes on the Sun \citep{Sheeley1976}, however the origins of the slow solar wind are still unclear \citep{Abbo2016}. There are two types of theory that seek to explain the origin of the slow solar wind. The first type states that slow wind originates in open magnetic field regions, where the field at the base of the corona is permanently connected to the heliosphere. These regions can either be the edge of coronal holes \citep{Levine1977, Wang1991} or the edge of active regions \citep{Harra2008, Harra2017, He2010}. To produce slow wind speeds they must be areas where the expansion of the magnetic field between the base of the corona and the heliosphere is largest \citep{Leer1980, Cranmer2007a, Wang2010} or where field line deviations from the radial direction at the coronal base are largest \citep{Pinto2016}. This theory predicts that the release of slow solar wind plasma into the heliosphere is a continuous steady state process.

The second set of theories are based on slow wind that starts in closed field regions, where the coronal magnetic field is not connected to the heliosphere. In these theories, interchange magnetic reconnection between adjacent open and closed field lines allows plasma that starts on closed field lines to access the open field lines and escape into the heliosphere \citep{Crooker2002, Crooker2004}. This can either happen throughout the closed field regions \citep{Fisk2001, Fisk2003, Fisk2008} or only at the open-closed boundary \citep{Antiochos2011, Antiochos2012, Higginson2017a}. These theories predict that the release of plasma is a transient process, occurring only when interchange reconnection takes place.

One transient component of the slow solar wind that has been observed both remotely and in-situ are discrete structures with enhanced number density. They were first discovered in remote sensing measurements by \cite{SheeleyJr.1997} using white light images, formed when sunlight Thompson scatters off electrons in the solar wind. When viewed in white light the structures are associated with the top of streamers, which are loops of closed magnetic field lines that separate open field regions of opposite polarity \citep{Rouillard2010a, Rouillard2011, Viall2015}. The two dimensional extension of the polarity inversion into the heliosphere forms the heliospheric current sheet (HCS), which starts above the tips of streamers, and is surrounded by dense material called the heliospheric plasma sheet \citep{Winterhalter1994}. Further away from the Sun dense structures form part of the material that surrounds the HCS \citep{Wang1998}. Once in the heliosphere their trajectories match what would be expected if they are periodically ejected from a fixed point co-rotating with the surface of the Sun \citep{Sheeley2010} and they are frozen in to and travel at the same speed as the surrounding solar wind \citep{Viall2010a}. All of these features agree qualitatively with interchange reconnection taking place near the edge of streamers, releasing dense plasma to form the structures which subsequently surround the HCS as they travel outwards into the heliosphere \citep{Sanchez-Diaz2017a}.

Remote observations of structures in the solar wind have a number of limitations. To be detectable the structures must have a large enough density relative to the background solar wind to scatter enough sunlight, have a large enough size to be resolvable by a finite resolution imaging instrument, and images must be exposed on short enough timescales so any measured intensity enhancements do not ``wash out'' due to smearing across multiple pixels. Some success has been had tracking individual structures up to 1 AU, however this often relies on the structures being compressed to increase their number density above a theoretically expected $1/r^{2}$ decrease \citep{Sheeley2010, DeForest2011, Howard2012}.

In-situ measurements do not suffer from these restrictions but instead are limited by the location of the observing spacecraft. Studies using data taken at 1 AU have shown individual cases of proton number density enhancements at the HCS that were accompanied by similar enhancements in Helium, Oxygen, and Carbon number densities \citep{Kepko2016}. The C$^{6+}$/C$^{5+}$ ratio was used to infer the temperature of the corona where each packet of solar wind originated. The derived coronal temperatures were very variable on timescales of the same order as the structure sizes, indicating that the release of slow solar wind was not a steady state process, but there was no clear correlation between the location of enhanced proton number density and inferred coronal temperatures. This meant the solar origin of the structures could not be unambiguously identified.

Previous studies, both remote sensing and in-situ, give an overview of structures generated near coronal streamers which form part of the material that surrounds the HCS in the heliosphere. In this paper we present the first in-situ statistical study of number density structures located in the slow solar wind, but away from in-situ crossings of the HCS. When observing the solar wind in-situ at 1 AU it is difficult to determine whether observations are truly a reflection of source properties, or are as a result of dynamical interactions as the solar wind travels radially outwards. For this reason we use data from the the Helios mission taken between 0.3 AU and 0.5 AU, described in sect. \ref{sec:data}. In sect. \ref{sec:single blob} we present observations of a single representative structure, and in sect. \ref{sec:stats} a survey of 140 identified structures summarises the plasma properties consistent across all structures. Finally, in sect. \ref{sec:discussion} we discuss the implications of these observations for the sources and formation of the slow solar wind.
\section{Data}
\label{sec:data}
\subsection{Ion distribution function processing}
This study uses data from the Helios mission \citep{PORSCHE1977}. Helios 1 and Helios 2 travel in elliptical heliocentric orbits between 0.3 AU and 1 AU, and were operational between 1974 - 1985 and 1976 - 1980 respectively. Each spacecraft has an electrostatic analyser for measuring ion distribution functions, which measured a full 3D distribution every 40.5 seconds  \citep{Schwenn1975}. The 3D distribution was then transmitted to the Earth; due to telemetry constraints the data is often not available at the full 40.5 second cadence. We have fitted the original ion distribution functions with a single bi-Maxwellian to extract the number density ($n_{p}$), bulk velocity ($\mathbf{v}_{p}$), and magnetic field parallel ($T_{p\parallel}$) and perpendicular ($T_{p\perp}$) temperatures of the proton core population. The fitting was done to each distribution using a least squares method that minimised the cost function
\begin{equation}
	C = \sum \left [f\left (n_{p}, \mathbf{v}_{p}, T_{p\parallel}, T_{p\perp} \right ) - f_{data} \right ]^{2}
\end{equation}
where $f\left (n_{p}, \mathbf{v}_{p}, T_{p\parallel}, T_{p\perp} \right )$ is the fitted bi-Maxwellian distribution function, $f_{data}$ is the measured distribution function, and the sum is taken over all points in $f_{data}$. Empirically the fitting process did a good job of fitting to only the proton core population, and was not affected by either the faster proton beam or alpha particle population that are sometimes present in the solar wind. An example fit can be viewed in figure 1 of the guide to the dataset \citep{Stansby2017c}. Along with the proton parameters, simultaneous magnetic field values were calculated as the average magnetic field during the time it took for each ion distribution function to be measured. Where available 4 samples/s data from the E2 fluxgate magnetometer were used \citep{Musmann1975}. When this was not the case 6 second cadence data from the E3 fluxgate magnetometer were used \citep{Scearce1975}.

The fitting process was applied to all the original ion distribution functions available from both Helios 1 and Helios 2. The code used to generate the dataset is openly available, making the fitted parameters easily reproducible. The source code and dataset, along with a more detailed description of the fitting procedure is available online \citep{Stansby2017c}.

From the fitted parameters the total proton temperature was calculated as
$
	T_{p} = \left (2T_{p\perp} + T_{p\parallel} \right ) / 3
$,
from which the proton thermal pressure is given by
$
	p_{p} = n_{p} k_{B} T_{p}
$.
The the proton plasma beta is defined as
$
	\beta_{p} = p_{p} / p_{mag}
$
and the magnetic pressure given by
$
	p_{mag} = \left | \mathbf{B} \right |^{2} / 2\mu_{0}
$

As well as protons, the other major constituents of the solar wind are alpha particles and electrons. Because corresponding electron and alpha number densities, velocities and tempereatures are not available it was not possible to calculate their contributions to the thermal pressure on a point by point basis. In slow solar wind alpha particle temperatures are between 0.2 - 1 MK \citep{Marsch1982c}, the alpha to proton number density ratio between 0.01 - 0.05 \citep{Kasper2007}, and electron temperatures lie in the range 0.2 - 0.4 MK \citep{Stverak2015}. Assuming plasma neutrality the proton, alpha, and electron number densities are related by $n_{p} + 2n_{\alpha} = n_{e}$. This relation along with the ranges of possible alpha and electron parameters allowed lower and upper bounds to be placed on the total pressure, defined as
$
	p_{tot} = \left (p_{p} + p_{\alpha} + p_{e} \right ) + p_{mag}
$.
In the results section values of $p_{tot}$ are presented as bounds or error bars between the minimum and maximum possible values.

\subsection{Number density structure selection}
\label{sec:structure selection}
From the entire Helios plasma data set a search was performed to find all the occurrences of number density structures. The search was limited to heliocentric distances ranging from 0.3 AU to 0.5 AU to minimise the time for stream-stream interactions to alter the local plasma properties. Structures were identified by eye in time series data as a clear sharp increase and subsequent decrease in proton number density accompanied by a similar signature in $\beta_{p}$. An example is shown in the bottom set of panels in fig. \ref{fig:single blob}. To distinguish structures from background noise they were required to last for more than 3 data points ($\sim$2 minutes).

Simultaneous increases in $n_{p}$ and $\beta_{p}$ are also properties of the heliospheric plasma sheet, which surrounds the HCS \citep{Winterhalter1994}. In order to clearly distinguish between the heliospheric plasma sheet and transient structures, any occurrences of the plasma sheet were avoided by ignoring enhanced number densities that surround the HCS. The HCS was identified by a change in sign of the radial magnetic field component ($B_{r}$) and coincident change in sign of the magnetic field-velocity correlation. The magnetic field-velocity correlation was used to distinguish between global polarity reversals and magnetic field foldbacks that only locally change the sign of $B_{r}$. This is similar to the common method of using solar wind electrons as a diagnostic \citep[e.g.][]{Owens2017}, but replaces the direction of electron strahl propagation with the direction of Alfv\'en wave propagation. This method assumes that Alfv\'en waves are always travelling along magnetic field lines away from the end connected to the Sun; this a safe assumption to make as very few examples of waves travelling towards the Sun have been reported \citep{Gosling2009a, He2015a}. Reconnection exhausts in the solar wind also exhibit the same increases in $n_{p}$ and $\beta_{p}$ \citep{Gosling2006}. To avoid selecting such exhausts any events which contained a smooth reversal in the magnetic field polarity were also discarded.

In total 140 structures were identified. The observing spacecraft and times of each structure are available online \citep{Stansby2018}.
\section{Results}
\subsection{An example density structure}
\label{sec:single blob}

\begin{figure}
	\resizebox{\hsize}{!}{\includegraphics{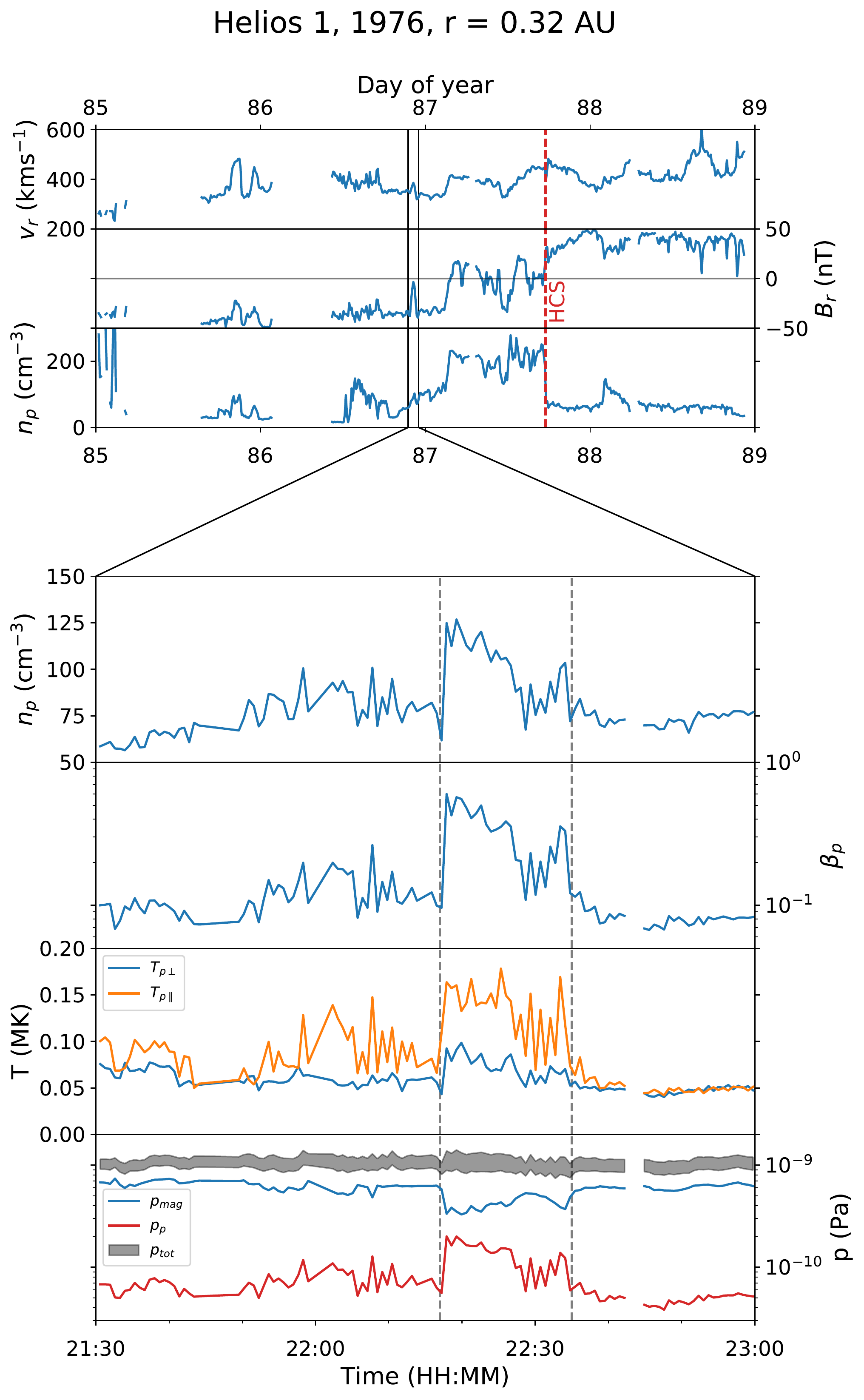}}
	\caption{An example number density structure. The top three panels show a 4 day overview of 10-minute averaged radial proton bulk speed ($v_{r}$, top axis), radial magnetic field ($B_{r}$, middle axis), and proton number density ($n_{p}$, bottom axis). The HCS crossing is marked with a red dashed line between days 87 and 88. The bottom 4 panels show a shorter 1.5 hour interval, with the structure present between the grey dashed lines. The top axis shows proton number density ($n_{p}$), and the second axis proton plasma beta ($\beta_{p}$). The third axis shows proton parallel temperature (orange, $T_{p \parallel}$) and proton perpendicular temperature (blue, $T_{p \perp}$). The fourth axis shows the proton thermal pressure (red, $p_{p}$), magnetic pressure (blue, $p_{mag}$), and total pressure (grey, $p_{tot}$). The range of values in the total pressure is due to uncertainty in the number density and temperature of electrons and alpha particles.}
	\label{fig:single blob}
\end{figure}

Figure \ref{fig:single blob} shows an example density structure, observed at 0.32 AU by Helios 1 on 28 March 1976. The top two panels show a 4 day overview of 10 minute averaged data, with the structure situated between the vertical black lines. The structure was located in slow solar wind ($v_{r} \approx$ 350 kms$^{-1}$) with a clearly defined negative magnetic polarity, around 18 hours before the in-situ HCS crossing.

The bottom set of panels shows a shorter time interval, with the structure beginning shortly after 22:15 and lasting around 20 minutes. It is visible as a proton number density enhancement, from a background level of 75~cm$^{-3}$ to a peak of 125~cm$^{-3}$. The background number density had a similar value before and after the structure. The relative density increase along with similar before and after density values are comparable to previous in-situ observations of structures at 1 AU \citep{Howard2012, Kepko2016}. Outside the structure $\beta_{p} \ll 1$ as is typical at 0.3 AU, whereas during the structure $\beta_{p}$ rose to 0.5.

Both of the proton temperatures were higher during the structure; the parallel temperature increased by a factor of 2 from 0.07 MK to 0.15 MK, and the perpendicular temperature underwent a smaller increase from 0.05 MK to 0.08 MK. This equated to a doubling in total temperature from $\sim$0.06 MK to $\sim$0.12 MK. The large increase in thermal pressure (caused by both increased number density and temperature) was offset by a similar decrease in magnetic pressure. Within uncertainty in electron and alpha particle pressures this resulted in no variation in the total pressure, meaning the structure was pressure balanced with its surroundings. The proton velocity was slightly higher inside the structure, rising from 330 km/s to 360 km/s (not shown). Because the structure was pressure balanced and travelled at a similar speed to the surrounding plasma, it is not expected to have undergone a significant interaction with the surrounding solar wind as it travelled away from the Sun.

\subsection{Statistical study}
\label{sec:stats}
For all structures identified using the method described in sect. \ref{sec:structure selection} average densities, velocities, temperatures, magnetic field strengths, and pressures inside each structure were computed, along with the same quantities averaged outside each structure. For quantities outside the structures an average of 3 data points ($\approx$ 2 minutes) before and 3 data points after the structure were taken.

\begin{figure}
	\resizebox{\hsize}{!}{\includegraphics{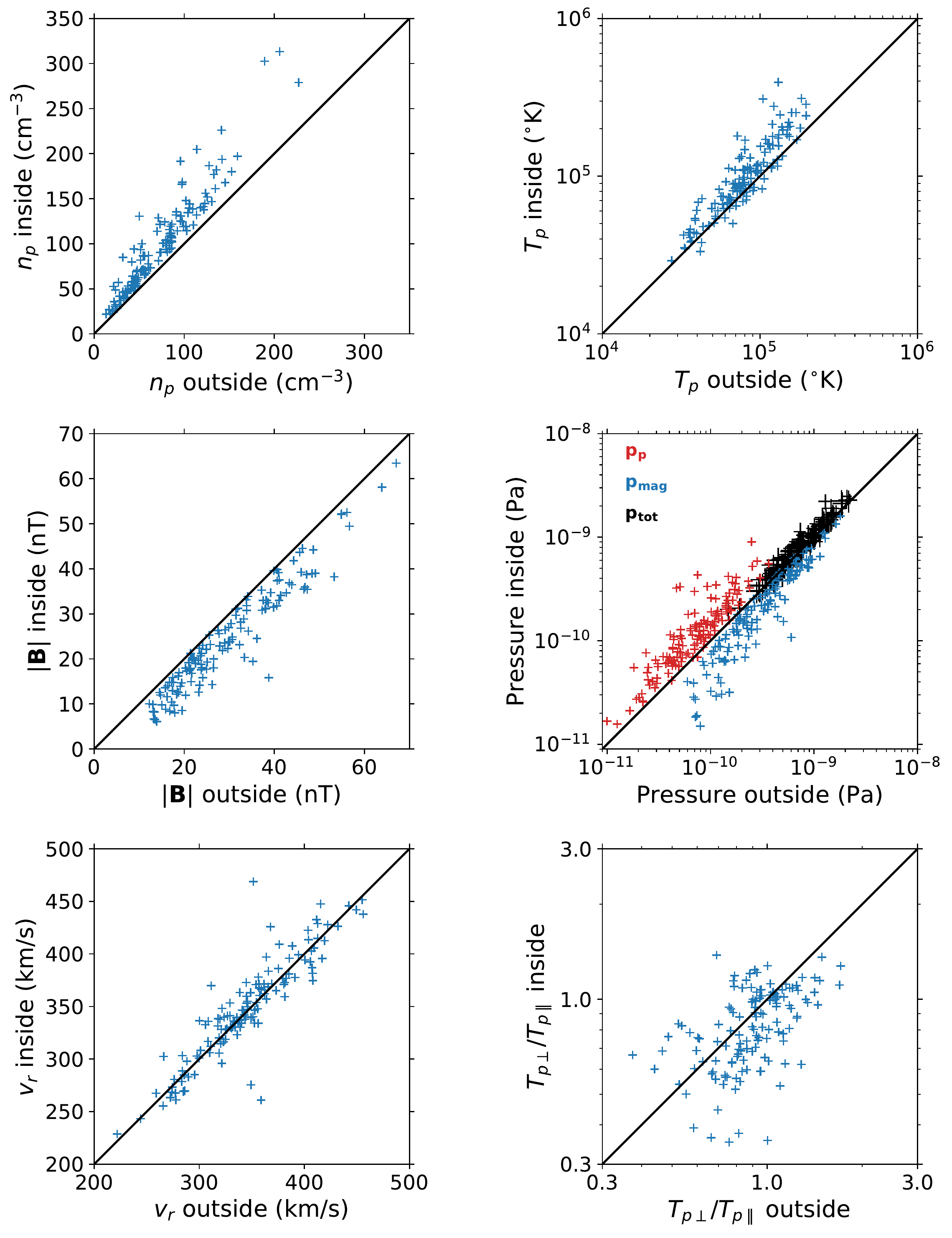}}
	\caption{Scatter plots of quantities inside the structures (y-axis) against quantities outside the structures (x-axis). One-to-one line are drawn in black on each plot. Points lying above these lines indicate a quantity larger inside a structure relative to outside and vice versa. Top left shows proton core number density ($n_{p}$) and top right total proton temperature ($T_{p}$). Middle left shows magnetic field magnitude ($\left | \mathbf{B} \right |$) and middle right magnetic (blue) proton thermal (red) and total (black) pressures. The total pressure crosses are error-bars in both directions due to uncertainty in the electron thermal pressure contribution to the total pressure. Bottom left shows radial proton velocity ($v_{r}$) and bottom right shows proton temperature anisotropy ($T_{p\perp} / T_{p\parallel}$).}
	\label{fig:scatters} 
\end{figure}

Figure \ref{fig:scatters} shows scatter plots of quantities inside the structures against average quantities outside. Each point in the plots represents a single structure. The top left panel shows that number densities were always higher in the structures, as originally selected for. The other five panels show quantities that were not visible during manual selection of the structures. Temperatures were generally larger inside the structures, with an upper limit of 2 times hotter than the surrounding plasma. This increase in temperature was not consistently partitioned between either the parallel or perpendicular temperatures, but the temperature anisotropy was often significantly different inside the structures. The magnetic field strength was lower inside the structures compared to outside. This indicates that thermal pressure increases caused by number density and temperature increases were accompanied by coincident magnetic pressure decreases. Thermal pressures were higher inside the structures, magnetic pressures lower, but the total pressure of the surrounding solar wind was similar to the total pressure inside the structures. This is consistent with the structures being pressure balanced with their surroundings. The magnetic strength decreases inside the structure were often accompanied by large changes in the individual field components. Variations in the components did not consistently happen in one direction, but the field magnitude and direction was consistently the same before and after each structure (not shown). Finally, the radial velocity of the structures was very similar to the surrounding solar wind.

\subsection{Viewing the structures in white light images}
Structures can be observed either in-situ or using remote sensing. In this section we link these two measurement techniques by using in-situ observations to predict which individual structures would be visible in white light images. The two structure properties that directly determine visibility in white light images are the fractional increase in electron number density and line of sight structure size. The fractional increase in number density was calculated as $\delta n / n_{0}$, where $n_{0}$ is the average proton number density outside the structure and $\delta n = \left \langle n \right \rangle - n_{0}$ is the average number density increase inside the structure. The available in-situ measurements are of proton number density, which does not necessarily equal the electron number density. The other major positive ion species in the solar wind is a population of alpha particles, therefore the electron density ratio $\delta n / n_{0}$ is only equal to the proton density ratio under the assumption that the proton to alpha particle ratio does not change across the structure. This may not be the case and in the slow solar wind values of the alpha particle abundance range between 1\% and 5\% \citep{Kasper2007}. This range of possible abundances translates to a maximum uncertainty of $\pm 8\%$ in the electron number density ratio.

When structures are viewed remotely the line of sight structure size is perpendicular to the radial direction of propagation. Only the flow parallel size is available from single spacecraft in-situ measurements, so this was used as a proxy for the line of sight size. This assumes that the structures have an aspect ratio of 1:1. The size of each structure was calculated as $l = t \left | \left \langle \mathbf{v}_{p} \right \rangle \right |$ where $l$ is the total advected size, $t$ is the structure duration, and $\left \langle  \mathbf{v}_{p} \right \rangle$ is the average proton velocity inside the structure. 

The visibility of the structures also depends on their location relative to the observing instrument. The strongest white light signal is achieved when a structure is located on the Thompson sphere, which has its centre half way along the observer-Sun line and passes through the observer and Sun \citep{Howard2009}. In appendix \ref{app:white light} it is shown that for a structure located on the Thompson sphere the fractional increase in intensity when viewing the structure is given by
\begin{equation}
	\frac{\delta I}{I_{0}} \approx 0.64 \frac{l}{r_{0}} \frac{\delta n}{n_{0}}
	\label{eq:intensity change}
\end{equation}
where $\delta I$ is the increase in white light intensity received when looking at a structure, $I_{0}$ is the background intensity when the structure is not present, and $r_{0}$ is the distance from the centre of the Sun to the structure. $\delta I / I_{0}$ is used directly when detecting number density structures in white light images.

\begin{figure}
	\resizebox{\hsize}{!}{\includegraphics{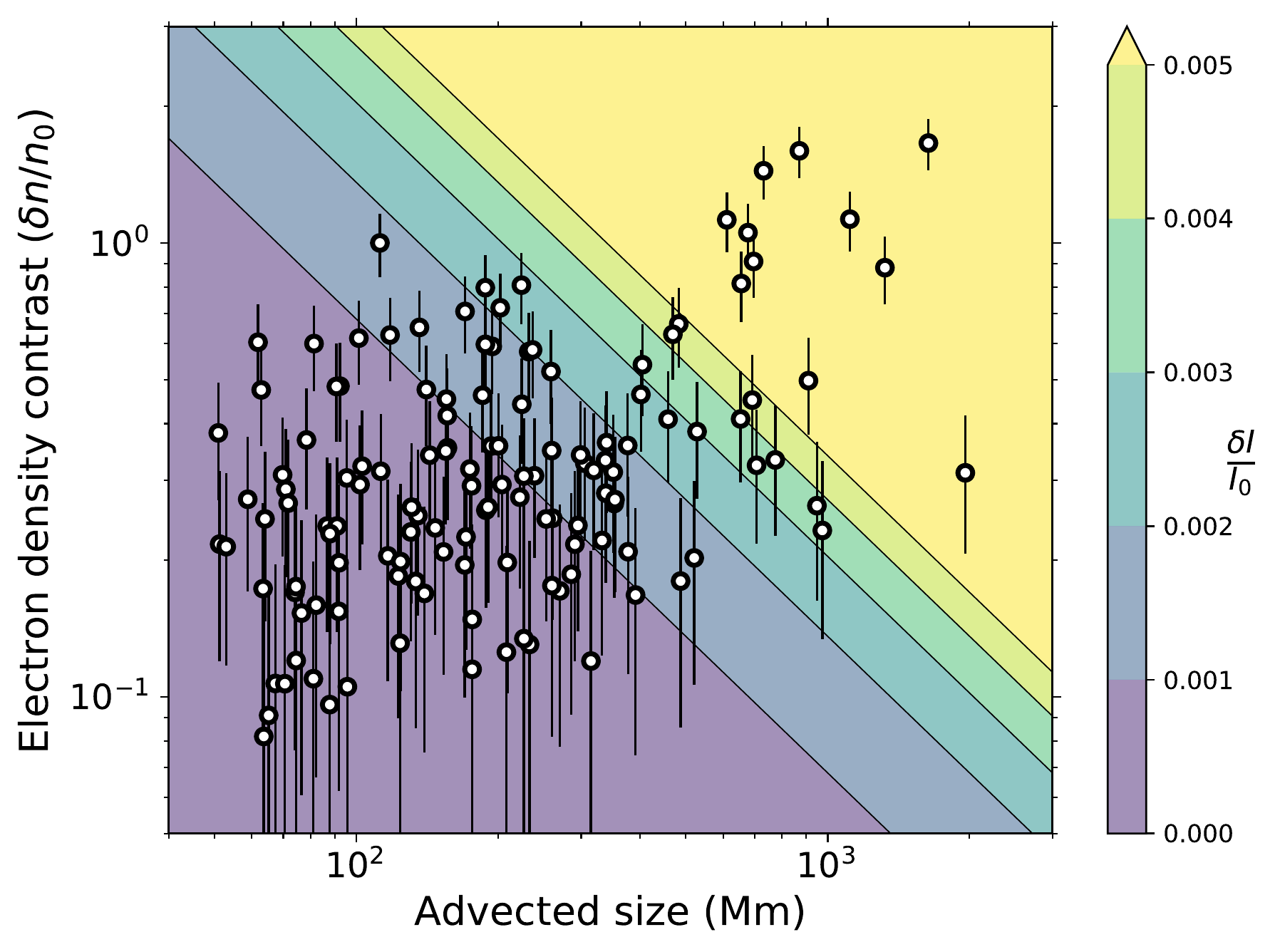}}
	\caption{Scatter plot of electron density contrast (y-axis) against advected structure size (x-axis) . Uncertainties in electron density contrast are due to not knowing the variation of the alpha particle abundance within the structures. Background shading shows contours of equation \ref{eq:intensity change} with $r_{0} = 0.3$ AU, which gives the fractional change in intensity a white-light instrument at 1 AU would see when viewing the blobs. The fractional intensity change values assume zero velocity smear across the observing instrument.}
	\label{fig:size} 
\end{figure}

Figure \ref{fig:size} shows a scatter plot of fractional number density enhancement against structure size, with constant contours of $\delta I / I_{0}$ from equation $\ref{eq:intensity change}$ in the background. The structures occur on scales ranging from 50 Mm to 2000 Mm, with no one characteristic scale size. The lower bound of 50 Mm is an instrumental limit caused by the 40.5 second cadence of plasma measurements available from Helios, along with the detection requirement that the structure duration spans at least three consecutive data points. The density contrast of the structures ranges from 10\% above background levels to 200\% above background levels with no single characteristic increase. Again the lower bound of 10\% is likely to be an artificial limit set by the minimum increase needed to manually distinguish a structure from noise in the proton number density measurements.

Equation \ref{eq:intensity change} assumes that a structure stays in a single pixel location during the time the image is exposed. In reality this is not the case, and $\delta I / I_{0}$ is reduced by a factor of the number of pixels the structure travels through during a single exposure. This is known as ``velocity smear''. For typical slow solar wind speeds of 300 km/s the instrumental velocity smear on Stereo HI-A reduces visibility by a factor of 10 \citep{Howard2008, Viall2010a}. Typical values for previously observed structures were in the range $\delta I / I_{0}$ = $2 \times10 ^{-4}$ - 10$^{-3}$ \citep{Viall2010a}, where the lower bound is due to the instrumental noise limit. Taking into account the velocity smear, values calculated using equation \ref{eq:intensity change} must satisfy $ \delta I / I_{0} > 2 \times 10^{-3}$ to be visible. 70\% of structures in fig. \ref{fig:size} do not meet this criteria, and thus would not be observable in Stereo HI-A images.

\section{Discussion}
\label{sec:discussion}
In the previous section 140 discrete areas of enhanced number density coinciding with enhanced proton beta were identified (for an example see fig. \ref{fig:single blob}). It was then shown that compared to the surrounding plasma these regions were hotter, had lower magnetic pressure and higher thermal pressure, but remained pressure balanced, and travelled at similar speeds to the surrounding solar wind (fig. \ref{fig:scatters}). The structures here had no characteristic size or density increase, and 70\% of them would not be visible in remote images taken at 1 AU (fig. \ref{fig:size}).

These observed features agree with previous individual in-situ measurements of pressure balanced structures in the inner heliosphere and at 1 AU \citep[e.g.][]{Burlaga1970, 1990AnGeo...8..713T, Yao2012} and this work has expanded on these studies to provide a statistical survey of individual structures on a wide range of scales. Note that because the magnetic pressure decreased and thermal pressure increased within the structures, all of them had a high plasma beta compared to their surroundings; this is in contrast to other pressure balanced structures with a low plasma beta which are also commonly observed in the solar wind, and often contain magnetic flux ropes \citep{Moldwin2000, Cartwright2008, Kilpua2012, Yu2014b, Feng2015}. Simulations have recently shown that pressure balanced structures can develop in a turbulent plasma such as the solar wind \citep{Yang2017a}. Dense structures have been tracked all the way from the surface of the Sun to 0.5 AU in white light images \citep{Sheeley2010}, which suggests that the most visible structures analysed here have a solar origin as opposed to developing as part of the solar wind turbulence. This does not rule out the possibility that the 70\% of structures that would not have been visible in white light images were formed in-transit as part of the solar wind turbulence.

Pressure balance across the structures means that no net force was acting on the them to alter their shape or size at the time of observation. We therefore predict that the only way the structures changed shape as they propagated away from the Sun was by undergoing a simple 1/r expansion in the two directions perpendicular to the radial direction of flow. In addition this validates the naming of the areas as `structures' which simply travel with the bulk solar wind flow, and agrees with remote sensing observations that came to a similar conclusion \citep{Viall2010a}. In contrast to remote sensing observations, the range of observed structure sizes here (50 Mm to 2000 Mm) contains scales much lower than the previous remote sensing lower limit of 1000 Mm \citep{Viall2010a}. We have shown quantitively that this is because the smallest structures are too small to be resolved by white light images taken from 1 AU.

The transient nature of the structures suggests that their release into the heliosphere was not continuous, and that (unlike coronal hole plasma) they did not originate on field lines that have a continuous connection from the corona to the solar wind. This conclusion leads to the interpretation that the structures were produced by interchange reconnection allowing hot, dense plasma originating on initially closed field lines to escape on to open field lines that map to the heliosphere (the mechanism sketched in \cite{Crooker2004}, fig. 5). The wide range of structure sizes implies that interchange reconnection was happening on a large range of physical length scales. Additionally all of the structures were observed in slow solar wind ($\left | \mathbf{v}_{p} \right | < 500$ km/s).

In addition to streamers, closed field regions can also take the form of pseudo-streamers which are pairs of closed loops that separate open field lines of the same polarity, and therefore do not have an associated current sheet \citep{Wang2007}. The structures observed here were all located away from in-situ crossings of HCS (by selection), which means that if interchange reconnection produced these structures it may have happened near pseudo-streamers as opposed to streamers. However, with only single spacecraft in-situ measurements and no remote sensing observations it is hard to tell how close to the HCS the spacecraft was at any given time. Simulations of interchange reconnection at the edges of streamers have produced blobs with high number density and high-$\beta$ which surround the HCS \citep{Higginson2017}. Further simulations could investigate whether structures can also be produced via interchange reconnection at pseudo-streamers, and any similarities and differences that could be used to distinguish between streamer and pseudo-streamer produced structures using in-situ measurements.

Finally, despite a search of the entire Helios mission only 140 examples of density structures away from crossings of the HCS were identified, constituting 40 hours of the 3,917 hours of slow solar wind ($\left | \mathbf{v}_{p} \right | <$ 500 km/s) Helios measured at distances $< 0.5$ AU. The structures presented here are rare, and did not make up the bulk of the slow solar wind. Similar blobs are frequently observed above streamers and surrounding the HCS, which we have not investigated here \citep{Sheeley2010, Rouillard2010a, Rouillard2010, Rouillard2011, Viall2015}, and \cite{Sanchez-Diaz2017a} estimate that this material forms 15\% of the slow solar wind. This means that even if all of the structures presented in this paper were generated at the edges of pseudo streamers, structures generated near pseudo-streamers are much rarer than those generated near streamers.

\section{Conclusion}
We have presented the first in-situ statistical study of transient number density structures in the inner heliosphere. The structures had a high plasma beta, were hotter than their surroundings, pressure balanced, and occurred in the slow solar wind but away from in-situ crossings of the HCS. These results suggest that away from the heliospheric plasma sheet, a small fraction of slow solar wind is produced by interchange reconnection releasing hot, dense plasma from closed field lines on to open field lines and out into the heliosphere.

In the near future Solar Orbiter will provide the first solar wind composition measurements in the inner heliosphere, allowing diagnosis of the coronal source of the structures. In addition Parker Solar Probe will provide the first in-situ observations of the solar wind at 10 solar radii, and will measure the structures shortly after they are released from the Sun. In combination with remote imaging of the structures, these new measurements will unambiguously allow us to identify the solar source of the structures.

\begin{acknowledgements}
The authors acknowledge useful discussions with Chadi Salem, Lorenzo Matteini, and Alexis Roulliard, and thank the anonymous reviewer for comments that significantly improved discussion of the results. D. Stansby is supported by STFC studentship ST/N504336/1. T. S. Horbury is supported by STFC grant ST/N000692/1. The Helios plasma dataset used here and described in sect. \ref{sec:data} is available online, along with the source code used to generate the dataset \citep{Stansby2017c}. Data was retrieved using HelioPy v0.4.2 \cite{Stansby2018a}, processed using astropy v3.0 \citep{2018arXiv180102634T}, and figures were produced using Matplotlib v2.1.1 \citep{Hunter2007, Droettboom2017a}.
\end{acknowledgements}

\begin{appendix}
\section{White light intensity signal of a single structure}
\label{app:white light}
The total intensity received by a white light detector is \citep{Howard2009}
\begin{equation}
	I = \int_{0}^{\infty} z^{2} n_{e} \left ( z \right ) G \left ( z \right ) dz
\end{equation}
where the integral is taken along the line of sight, $n_{e}$ is the electron number density, and $G \left ( z \right )$ is the scattering expression. $I_{0}$ is defined as the value of the integral when $n_{e} = n_{0} \propto 1/r^{2}$, where $r$ is radial distance from the centre of the Sun (i.e.\ an unperturbed number density profile for the solar wind). Because the integral is linear in $n_{e}$, the total intensity when a structure is added to the background profile is
\begin{equation}
	I = I_{0} + \int_{z_{0} - l / 2}^{z_{0} + l / 2} z^{2}  \left [ n_{e}\left ( z \right ) - n_{0}\left ( z \right ) \right ]  G \left ( z \right ) dz
\end{equation}
where the structure is centred at $z_{0}$ and has a line-of-sight length $l$. Under the assumption that the value of the integrand is constant over the structure this further reduces to
\begin{equation}
	I = I_{0} + l z_{0}^{2}\left [ n_{e}\left ( z_{0} \right ) - n_{0}\left ( z_{0} \right ) \right ]  G \left ( z_{0} \right )
\end{equation}
The change in intensity relative to the background intensity is then given by
\begin{equation}
	\frac{\delta I}{I_{0}}= \left [\frac{z_{0}^{2} G \left ( z_{0} \right ) n_{0} \left ( z_{0} \right )}{I_{0}} \right ] l\ \frac{\delta n}{n_{0}}
\end{equation}
where $\delta I = I - I_{0}$ and $\delta n = n_{e} - n_{0}$. The term in square brackets has units of inverse distance. It depends explicitly on the structure location ($z_{0}$) and implicitly on the observer location and look direction. It takes its largest value when $z_{0}$ is located on the Thompson sphere \citep{Howard2009}. For an observer situated at 1 AU, small observation angles with respect to the Sun, and a structure located on the Thompson sphere, the expression in square brackets evaluates numerically to $\approx 0.64 / r_{0}$, where $r_{0}$ is the distance from the centre of the Sun to the number density structure. This gives the final expression
\begin{equation}
	\frac{\delta I}{I_{0}} \approx 0.64 \frac{l}{r_{0}} \frac{\delta n}{n_{0}}
	\label{eq:final intensity change}
\end{equation}
This equation gives the change in intensity relative to the background intensity if a structure is located at the most favourable position for viewing. Constant contours of $\delta I / I_{0}$ calculated using equation \ref{eq:final intensity change} are plotted on fig. \ref{fig:size} for the range of structures sizes and density increases observed in this paper.
\end{appendix}

\bibliographystyle{aa}
\bibliography{library}

\end{document}